**ARTICLE**

# Computer vision in tactical artificial intelligence art

**Dejan Grba***

Digital Art Program, Interdisciplinary Graduate Center, University of the Arts, Belgrade, Serbia

## Abstract

Artificial intelligence (AI) art comprises a spectrum of creative endeavors that emerge from and respond to the development of AI, the expansion of AI-powered economies, and their influence on culture and society. Within this repertoire, the relationship between the cognitive value of human vision and the wide application range of computer vision (CV) technologies opens a sizeable space for exploring the problematic sociopolitical aspects of automated inference and decision-making in modern AI. This paper examines art practices critically engaged with the notions and protocols of CV. After identifying and contextualizing CV-related tactical AI art, we discuss the features of exemplar artworks in four interrelated subject areas. Their topical overlaps, common critical points, and shared pitfalls plot a wider landscape of tactical AI art, allowing us to detect factors that affect its poetic cogency, social responsibility, and political impact, some of which are rooted in the theoretical premises of digital art activism. Along these lines, the paper outlines the routes for addressing the challenges and advancing the field.

## 1. Introduction

Contemporary artificial intelligence (AI) art is a diverse set of experimental, exploratory, and mainstream practices that utilize, contextualize, and mediate the phenomenology of AI, with a history dating back to the 1970s and earlier. Its topics, concepts, methodologies, and presentational formats capture and reflect various AI research and application trends.[1,2]

Marked by a significant concentration of wealth and power, AI's rapid industrialization brings about wide-reaching economic and political changes that occasionally prove controversial.[3] Notably, modern machine learning (ML) technologies increase the extent and intricacy of corporate or state strategies that combine digital datafication with statistical reductionism for profiteering or social engineering, often amplifying their undesirable effects.[4,5] In a critical discourse that traces and questions the constitution of social reality through modern AI's technical logic and business practices, artists play a significant role by deconstructing, assessing, appropriating, and repurposing AI technologies to uncover or underline AI development's existential and epistemic issues and the economic and sociopolitical consequences of AI capitalism. They continue the heterogeneous flux of tactical media practices that have energized art and culture since the late 20th century with hybrid forms of esthetic or critical interventions into the neoliberal condition's technical, political, economic, and cultural layers. Alternative names for tactical media art are digital activism, critical media, critical hacking, hacktivism, and culture jamming, although these terms sometimes also refer to specific genres. For a concise introduction to the field of tactical media, see Raley.[6]

Largely following tactical media art's theoretical principles and methodologies, artists who work with AI combine appropriation, revelation, provocation, and humor in approaching AI systems' operative modes or objectives. Successful works usually refrain from grandiose objectives, overt representation, and over-explanation in favor of subtle or sometimes intentionally ambivalent critique to let the audience actively learn about the controversial views and interests behind certain AI technologies and identify the animosities, struggles, inequalities, or injustices in their application frameworks. They emphasize presence, engagement, and response but occasionally constrain them to raise the awareness of limited actionability or immediate political impact. Notwithstanding their creative scenarios, tactical AI artworks are distinguished by their critically motivated intents and attitudes.[7]

Their topical range includes the sociopolitical issues of applied AI, such as inductive biases and prejudices or exerted inequities and injustices, the AI industry's exploitative practices, such as the aggregation of transnational labor for training and testing ML models, the systematic concealment of the extent, value, and implications of human creative endeavor in AI development, and workforce conditioning through manipulative interfaces, complacency, lack of protection, or precarity. The poetic values and social impact of tactical AI artworks are proportional to the breadth and depth of their creators' understanding of the sociopolitical, economic, and moral nuances dispersed across the abstract, technically convoluted, and functionally opaque realities of AI technologies and their applications. This proportionality makes tactical AI art conducive to studying how AI simultaneously reflects and influences contemporary relations, economies, and worldviews.

A diverse body of critical AI studies is useful for exploring this field. Scientists who investigate AI issues relevant to tactical AI art include Cathy O'Neil,[4] Gary Marcus and Ernest Davis,[8] Melanie Mitchell,[9] Erik Larson,[10] Matteo Pasquinelli,[11] Michael Kearns and Aaron Roth,[12] and Brian Christian.[13] Writers in the humanities' critical AI studies include Kate Crawford,[14] Simone Natale,[15] Vladan Joler and Matteo Pasquinelli,[16] and various authors contributing to the book *The Cultural Life of Machine Learning: An Incursion into Critical AI Studies*,[17] edited by Jonathan Roberge and Michael Castelle, and *Critical AI: A Field in Formation*,[18] edited by Rita Raley and Jennifer Rhee. Authors who explore AI art practices include Joanna Żylińska,[19] Martin Zeilinger,[20] and Dejan Grba.[2,7,21-23] While Żylińska, Zeilinger, and Grba address tactical artworks, their studies are not specifically about the critical practices centered around computer vision (CV),[1] which figures prominently in contemporary culture. Żylińska identifies the post-conceptual artistic treatments of the ultimate loss of meaning in the hyper-pragmatic development and application realms of machine vision systems and the practices with delegated labor that underscore the issues of AI's artificiality. Zeilinger focuses on artworks that simultaneously appropriate and oppose AI's capability of hacking agency, creativity, and ownership to challenge the integrity of these notions that underlie legal prescriptions about intellectual property and artistic uses of technology.

This paper complements their scopes by elaborating the critical perspective introduced in specific studies by Grba[7,22] to examine a diverse field of AI art's tactical engagement with the notions and protocols of CV. After providing the contextual overview of CV in computational art, it explores the poetic features of artworks that utilize CV in different conceptual, technical, and presentational arrangements to question AI's influence on social, economic, and political relations. The focus is on works produced in the past two decades when the coupling of statistical ML with CV technologies has empowered the pervasive surveillance culture to penetrate more offline and online layers of everyday life. They are reviewed in four interlinked subject areas of the central section: sociotechnical issues, control and conditioning, biometric classification, and ethical and epistemic limits.[2] Topographic mapping of the CV-related tactical art requires a separate full-volume study, so this four-section lineup highlights the creative approaches to CV that represent tactical AI art's most salient directives. The closing discussion identifies several problems that affect tactical art's impact on modern AI and outlines potential routes for tackling the challenges and advancing the field's conceptual cogency, critical strength, and social value.

Placing the successes and challenges of CV-related tactical practices within a broader context defined by

---

1    CV is the research area of computer science that seeks solutions for object and feature detection and classification, facial and gesture recognition, scene segmentation, motion tracking, and other forms of image-based automation and human-computer interaction.[24]

2    All discussed works are listed in the reference list and well documented online, so their descriptions have been compacted to the topically most pertinent aspects. Details of additional (mentioned but not discussed) exemplars can be found online by querying artists' names and work titles.

    **2**    

modern AI's socio-technical regimen, we trace their interrelated topics and imbricated critical points that reveal the info-capitalist suppression of AI technologies' social rootage and filtering human benefits from making and leveraging them, underline human roles behind AI's performative power, and expose human interests that drive social conflicts fueled by applied AI. This appraisal foregrounds artworks' effectiveness in instigating apparent and practical (not hypothetical or speculative) changes in human relations, economy, and politics for a more just and livable society. It considers CV-related critical art as a valuable tributary to the contemporary AI debate, but its missed potentials and failures indicate tactical AI art's shared vulnerabilities whose improvements may require reassessing the premises in the field's theoretical foundations.

## 2. CV in tactical AI art

Art is an inherently technological realm of human activity, and optical tools are among its key topics and obsessions, especially in Western cultural tradition,[3] so artists' interest in machine vision is not surprising. The immediate pretext for AI art practices with CV was set by the surveillance art genre in the three closing decades of the 20th century. Surveillance art emerged as a reaction to the proliferation and steady improvements of visual technologies in the military, law enforcement, science, and business; the global spread of CCTV and video satellite networks; and the rising popularity of mediated everyday life in reality TV shows and on the Internet.[28] Reinforcing the post-structuralist[4] social critique, these changes motivated artists to probe the burgeoning relationship between power and new vision technologies. They took various pathways using photography, film, video, installation, performance, and the Internet to deconstruct and question voyeurism, indifference, control, reliability, accuracy, and other features and aspects of machine vision in the surveillance culture.[30-32] Although surveillance art had variable (arguably negligible) success in instigating noticeable sociopolitical changes, it was generally well received by the audience and academia and influenced the art world.

Myron Krueger's interactive installation *Videoplace* (developed between 1969 and 1975)[33] pioneered the use of CV in the mid-1970s, but this technology figured in relatively few surveillance artworks before the 2000s because it was difficult to access and use for most artists and the contemporaneous surveillance culture was dominated by analog imaging technologies. One of the noteworthy exceptions is *Suicide Box* (1996)[34] by the Bureau of Inverse Technology (Natalie Jeremijenko and Kate Rich), which tooled the CV for a provocative social critique. With a motion-sensitive video camera setup that recorded only upon the detection of vertical motion, the artists covered the Golden Gate Bridge in San Francisco for 100 days to capture suicide jumpers. Finalized as an edited video of 17 registered incidents, the work stirred controversy at the 1997 Whitney Biennial.[35] Other artists who worked with CV in the 1990s include Simon Penny, Toshio Iwai, Christa Sommerer and Laurent Mignonneau, Camille Utterback and Romy Achituv, Scott Snibbe, and Marie Sester.

The release of versatile CV programming libraries, such as OpenCV and Dlib in the early 2000s, and the successes of data-driven ML technologies, pushed the limits of digital image processing and gave artists more freedom and flexibility to work with object- or feature-recognition data extracted from pictures, videos, and live camera feeds. The following diversification of generative and interactive experiments that explored perception, behavior, and cognition by interfacing humans with ML is illustrated by a string of cogent, technically accomplished projects. For instance, in the installation by Golan Levin and Greg Baltus *Opto-Isolator* (2007),[36] a wall-mounted box hosts a single mechatronic eye in the center and a camera pinhole on the side. The box surface is reflexive so visitors can see themselves while interacting with the eye, whose uncanny feedback repertoire includes looking the viewer directly in the eye, moving as if intently studying the viewer's face, looking away coyly if it is stared at for too long, or blinking precisely one second after the viewer (Figure 1). This is one of many artworks that use anthropomorphic premises about machine vision to create the effect of inverted art spectatorship by asking: "What if artworks could know how we were looking at them? And, given this knowledge, how might they respond to us?"[36]

Conversely and equally important, Seiko Mikami's installation *Desire of Codes* (2010)[37] underlines the abstract character of machinic perception. It consists of 90 wall-mounted mechatronic devices, six ceiling-suspended robotic arms equipped with cameras and various sensors, and a semi-spherical screen composed of 61 hexagonal cells. These interactive sections provide a complex, esthetically competent, although ultimately counterintuitive experience for the visitors to "investigate the forms of human corporeity and desire facilitated by surveillance technology and network society."[37] Related examples include Random International's installation

---

3   See, for example, Kubovy,[25] STOA,[26] and Witcombe.[27]

4   Michel Foucault's book *Discipline and Punish* (first published in 1975)[29] has been one of the influential theoretical sources for visual arts. It analyzes the changes in Western penal and social normative/disciplinary mechanisms (in schools, hospitals, and the military) that the new technological powers of corporeal control introduced during the modern age.

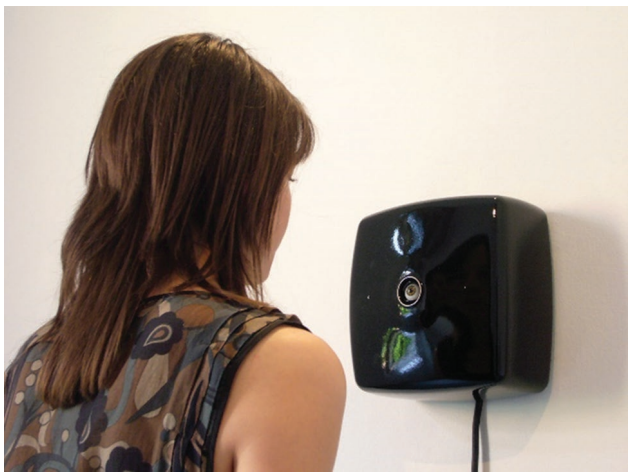

**Figure 1.** Golan Levin and Greg Baltus, *Opto-Isolator* (2007). Photograph courtesy of the artists

*Audience* (2008) and Morgan Rauscher's installation *Zeugen* (2009 – 2010).

The primary roles of CV in these works are to emphasize – through imitation or alienation – the psycho-social aspects of observation and gaze and the datafication of the human body and existence. They also raise the questions of spectatorship, the "ownership" of visual experience, the nature of spectacle, and the tensions between passive and active observation. Most of these works include critical overtones about automated surveillance, but their most pronounced poetic traits do not necessarily target the specific sociotechnical contexts of AI-influenced culture.

Modern CV contributes to AI's impact on the globalized infrastructures of industry, commerce, communication, and entertainment by funneling digital emulations of cognitively most beneficial human sense (sight) through complex but often flawed ML filters. Tactical AI artworks with CV address the practical, political, and ethical aspects of this confluence, which has allowed corporate and state sectors to expand in scope and reach new levels of sophistication and efficacy in supervision, data collection, biometric profiling, behavioral tracking and prediction, micro-targeting, and social engineering.[4,38] By creating technical arrangements that imitate, deviate from, or conflict with human visual perception, tactical AI artworks question, criticize, and subvert the roles of CV and the concept of seeing-as-knowing in these new power mechanisms. They use CV to point out the problematic and undesirable application consequences of ML architectures and indicate the broader epistemic, existential, and sociopolitical issues of modern AI. To incite critical scrutiny, artists modify, repurpose, or recontextualize CV technologies outside their established operative modes and domains. They do not always follow openly activist agendas but employ subtle or covert micro-politics of disruption, intervention, and education. For revelatory purposes, they occasionally combine humor (irony or sarcasm) with provocation and take idiosyncratic or ambivalent positions toward the issues they address. Sometimes, however, their missed potentials and failures indicate factors that affect tactical (AI) art's poetic cogency, social responsibility, and political impact. Here are some examples.

### 2.1. Sociotechnical issues

Many concepts, themes, and methodologies of recent CV-related AI art were introduced by tactical works that combined repurposed CV technologies with pattern recognition and natural language processing (NLP) in the 2000s and early 2010s. For example, Christian Moeller's *Cheese* (2003)[39] is one of the early artworks that used facial attribute classification for social critique. Moeller hired six actresses to hold a smile in front of a video camera for as long as they could (up to 90 min). Their facial expressions were scrutinized in real-time by an emotion recognition system, and whenever their grimace went below a certain parametric threshold of "happiness," an alarm alerted them to show more "sincerity." The resulting installation comprises video displays documenting each actress' effort together with a vertical graph that transitions from red to green according to her current "sincerity" level (Figure 2). *Cheese* expounds the hypocrisies of sociotechnical normalization of individual behavior and presentation.[5] By paying actresses to perform an ultimately absurd type of work regardless of their personal appraisal of its artistic purpose, Moeller also nods toward Santiago Sierra's ambivalent critique of labor and power disparities in performative setups, where participants (often from underprivileged communities) are hired for absurd, menial, or abusive physical tasks.[40]

Another landmark work, Paolo Cirio and Alessandro Ludovico's *Face to Facebook* (2010),[41] revealed the application of automated facial recognition in social media by hijacking and intersecting it with uncertainty and arbitrariness to make a multifaceted critique of AI-powered platform capitalism.[6] The artists created software bots that ran pattern recognition, CV, and NLP routines over online protocols to harvest one million Facebook profiles, filter out 250,000 profile photographs, custom-tag their facial expressions, and post them as new profiles on a fictitious dating website called *Lovely Faces*. In 5 days while *Lovely Faces* was online, the artists received several letters from Facebook's lawyers, 11 lawsuit warnings, and five death threats.

---

5    For a similar, albeit more abstractly formalized sociotechnical critique, David Rokeby's installations *Sorting Daemon* (2003) and *Gathering* (2004) used human figure recognition and color sorting from a live video feed to express concerns about the application of automated systems for personal profiling in the war on terror.

6    This is the final part of the *Hacking Monopolism Trilogy* (2006 – 2010), including *Google Will Eat Itself* and *Amazon Noir*, realized in 2006 by Cirio, Ludovico, and ÜBERMORGEN.COM.

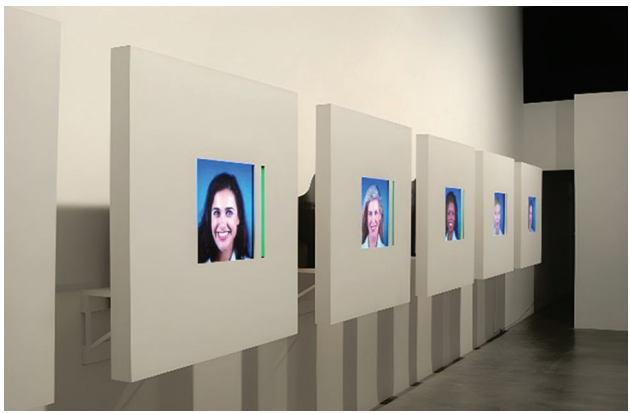

**Figure 2.** Christian Moeller, *Cheese* (2003). Installation view. Photograph courtesy of the artist

Some artists repurpose CV within the broader (international or electoral) contexts of politics. For instance, Kyle McDonald's *Sharing Faces* (2013 – 2014)[42] utilized CV with astounding economy to mediate empathy in a telematic interaction between people in two countries with historically uneasy relations. The work was installed for 8 months concurrently in Anyang, South Korea, and Yamaguchi, Japan. At each location, a camera/display setup emulated in real-time the current visitor's facial expressions and poses with their closest matches from a cumulative library of visitors' images captured in the opposite location. In *Smile to Vote* (2017 – 2019),[43] Alexander Peterhaensel elaborated on Moeller's concept in *Cheese* to make an engaging satire of technopolitics. Visitors of this interactive installation enter a voting booth where a camera captures their portraits and sends them to the *Smile to Vote* software. It uses OpenCV to "deduce" the visitor's political conviction and make the appropriate voting decision by comparing their facial features with a dataset of portraits pre-classified by unequivocal political orientation (party members and election candidates). The percentage of the visitor's physiognomic congruency with eligible parties is displayed as a bar chart on the screen, and a voting receipt is printed out for collection. By transferring the principle of Alibaba's "Smile to Pay" service[7] into the context of electoral politics, *Smile to Vote* highlights the influence of AI-powered biometric profiling on democratic processes, political convictions, self-determination, and privacy.

---

7    The Alibaba online shopping website has used the facial identification/payment system "Smile to Pay" since September 2017.

Targeting the use of CV for algorithmic decision-making in the human resources (HRs) business, Varvara Guljajeva, and Mar Cane's *Keep Smiling* (2022) "aims to scrutinize automated hiring, skills-testing, and dismissal systems [and] reflect on the implications of AI use in the hiring process."[44,45] It also exemplifies how artworks' critical impact can be compromised when the handling of sensitive issues excessively relies on the more compelling but unacknowledged precedents. Assuming the role of job interviewees, this online work's visitors interact with an AI hiring agent, whose requests to smile for an indefinite period they inevitably fail, resulting in the loss of the fictional job. The gameplay mixes the caricatural remediation of an online job interview with the insinuations of related AI issues, such as "fauxtomation."[46] However, it offers no helpful insights into the controversies surrounding the trend of "augmenting" HR business with AI; instead, the artists rely on a written introduction and conference papers about the work. Rather than disrupting the underlying logic of AI-powered HR, *Keep Smiling* merely subverts one of its mechanisms – automated emotion detection – within a simplified fictional setting, thus rendering their problematic social consequences somewhat easier to tolerate. Moreover, although the artists acknowledge the work's relationship with Moeller's *Cheese*, Peterhaensel's *Smile to Vote*, and Coralie Vogelaar's *Random String of Emotions* (2018, discussed below),[47] they omit mentioning its striking resemblance to Carrie Sijia Wang's *An Interview with ALEX* released in 2020,[48] whose central part places visitors in a 12-min job interview with a fictional AI-ran corporate HR system called ALEX (Figure 3). This earlier but more elaborately structured online project delivers a stronger narrative about the roles of AI in relational manipulation, job precarity, and gamification of labor.

Reflecting and extending their surveillance art predecessors' approach to video monitoring technologies, these works employ CV in experiential arrangements that underline visitors' presence and hint at the human roles in highlighting the consequences of the widespread AI application.

## 2.2. Control and conditioning

Mediated supervision and automated deliberation are prominent surveillance art topics, and the applications of CV in these control and command mechanisms provide plentiful pathways to tap into their (ab)uses in various sociopolitical contexts. For example, Ken Rinaldo's *Paparazzi Bots* (2009)[49] connected human-robot interaction with the capricious celebrity criteria in popular media and artist-superstar myths stoked by the art market ([Figure 4](#)). This work featured autonomous CV-controlled robots that roamed galleries and art events, picked up visitors who were smiling, took their snapshots, and uploaded them on the Internet and to the press services. By enticing visitors into an intentionally *manipulative* interaction (only those who smile get "rewarded" by being photographed), *Paparazzi Bots* extends its witty critique of the manipulative power of visual capture beyond the art world culture toward the social media business, which had already been heavily developing and applying AI to manage their services and users.

A conceptual relative to *Paparazzi Bots* and Random International's installation *Audience* (2008, mentioned above), Shinseungback Kimyonghun's *Nonfacial Mirror* (2013)[50] elegantly employs an inverse logic of interaction. It is a plinth that hosts a motorized bathroom mirror with a camera concealed behind the glass pane and OpenCV and Dlib face detection libraries running in the computer inside the plinth. The mirror persistently rotates away from the visitor if either one of the libraries detects a human face in the camera's view. Like its two predecessors, *Nonfacial Mirror* references earlier surveillance artworks, such as Peter Weibel's three-channel CCTV installation *Beobachtung der Beobachtung: Unbestimmtheit* (*Observing Observation: Uncertainty*, 1973),[51] whose cameras and monitors were arranged to prevent visitors from seeing their faces in any position they took inside the setup.

Besides highlighting the power of AI-driven surveillance systems to mediate public presence and self-perception, other works play with the nuances of sociocultural contextualization and technological conditioning of gaze by way of CV. For instance, Kenichi Okada and Naoaki Fujimoto's *Peeping Hole* (2010)[52] reinvigorated the well-established artistic exploration of eye-tracking techniques in a charming game of shared visual curiosity and transgression ([Figure 5](#)). As part of a group exhibition, this installation invites visitors to look through a peeping hole in one of the gallery walls. When a visitor starts looking at the beach scene photograph displayed in a chamber behind the wall (and only while looking), an overhead projector beams the photograph's circular area tracking their wandering focal interest on the wall's front surface above the peeper, who remains oblivious of other visitors' participation in their curiosity. Its friendly betrayal and subjugation of viewers' expectations in consuming art discreetly point to the wider issues of reliance on and manipulation of the audience's trust by both the artists and art institutions.

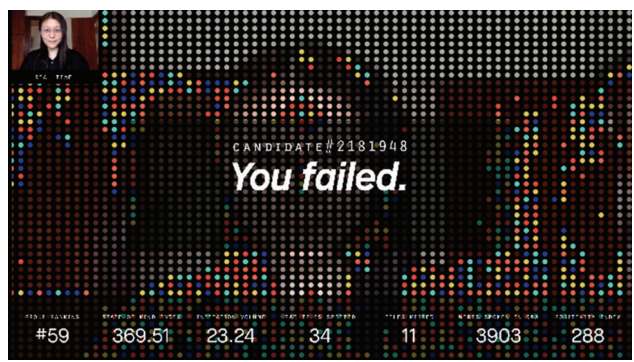

**Figure 3.** Carrie Sijia Wang, *An Interview with ALEX* (2020). Image courtesy of the artist

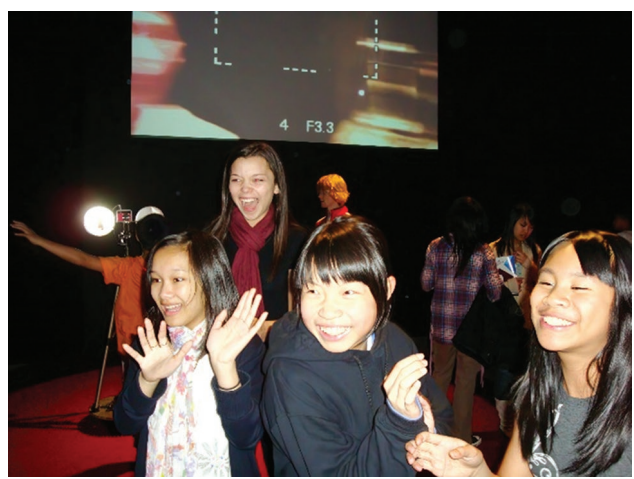

**Figure 4.** Ken Rinaldo, *Paparazzi Bots* (2009). Installation view at the Cultural Olympiad Digital Edition exhibition curated by Malcolm Levy, 2010 Winter Olympics, Vancouver, Canada. Photograph courtesy of the artist

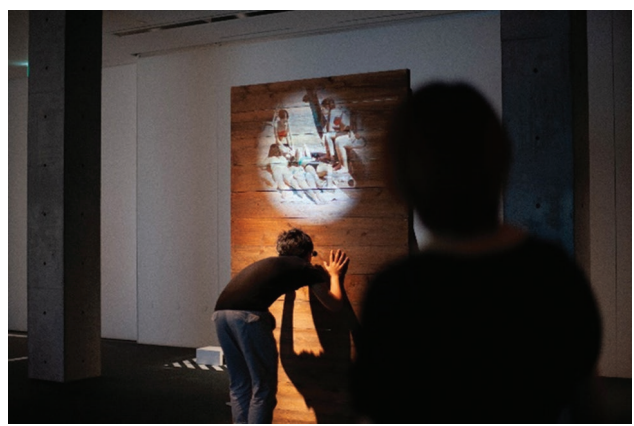

**Figure 5.** Kenichi Okada and Naoaki Fujimoto, *Peeping Hole* (2010). Installation view. Photograph: Kenichi Okada. Courtesy of the artists

Kyle McDonald and Matt Mets' *Blind Self Portrait* (2012)[53] took a different look at the trust-based (voluntary)

submission in technically mediated sociocultural relations. This face recognition setup draws linear portraits of the visitors, but to make it work, the sitter must keep their eyes closed while resting their pen-holding hand on a small platform that moves along the X/Y axes over a fixed piece of paper. By playfully turning visitors into the "mechanical parts" of a drawing apparatus and making them conscious of both their acquiescence and slight functional unreliability, *Blind Self Portrait* (knowingly or not) references William Anastasi's *Subway Drawings* from the 1960s.[54] Similarly, Patrick Tresset's *Human Studies* (since 2011)[55] is a series of posing sessions in which humans become passive sitters for portrait-drawing robots that imitate Tresset's visual style. However, Tresset started developing robots for *Human Studies* as creative prostheses to overcome his artistic block,[56] and the primary poetic feature of this non-tactical project is the novelty-induced anticipation of finished machinic drawings that relieves visitors' deference and typical modeling boredom.

It is easy to forget that the control and command "reign" of CV software routines is governed by a higher order of conditioning and routing algorithms. By intersecting this computational hierarchy facet with cultural heritage, artists can point out the constraints of AI-influenced society. For instance, in *Computers Watching Movies* (2013),[57] Ben Grosser combined the informative open-endedness of abstract forms with cumulative cultural experience to engage visitors in an imaginative guessing game (Figure 6). The work features six video sequences produced by the CV analysis of popular film sequences. Points and vectors of the CV program's "focal interest" (image locations assigned with higher weights) are animated as colored dots and lines on a white background (the processed film footage is not visible) and synchronized with the original film sound. The coupling of minimalist visuals with sonic guidance draws viewers into a series of comparisons between their culturally developed ways of seeing and CV software's "attention" logic, which has no historical, narrative, or emotional patterns.

Beyond their immediate topical concerns, these works connect with tactical art production that examines the roles of CV in cybernetic systems designed for labor maximization, profit extraction, and other questionable forms of societal control. For instance, Matt Richardson's *Descriptive Camera* (2012)[58] uses a novel form of taking snapshots to indicate the socioeconomic background of modern AI technologies. This camera has a lens, but instead of displaying the photographed image, it sends the image to an Amazon MTurker tasked with writing down and uploading its brief description, which the device prints out. It offers a revelatory, counterintuitive glimpse into the widespread exploitation of transnational echelons of underpaid workers for AI development, which has been identified as "fauxtomation,"[46] "sociotechnical blindness,"[59] "ghost work,"[60] and "human in the loop" complex.[61]

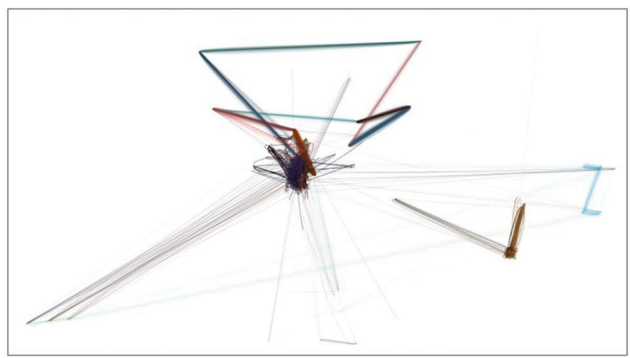

**Figure 6.** Ben Grosser, *Computers Watching Movies* (2013). Image courtesy of the artist

As the pragmatic functions of CV and other machine vision technologies, command and control are repurposed in these works to unveil the questionable or problematic aspects of AI in the economy, culture, and various layers of everyday life, which originate in the higher-level computational processes and human reasoning in their design and implementation.

## 2.3. Biometric classification

CV has been extensively integrated into a pipeline for extracting and combining different types of biometric data to build formal categorization models that can be used for contact filtering, individual identification, verification, profiling, and determination of rights, such as entry privileges, service access or modification, and pricing. Potential areas of application include state governments, military, intelligence, health and wellness, HR, commerce, and entertainment industries. Modern AI processes biometric data in statistical prediction frameworks whose reductiveness can be problematic even in hypothetical scenarios with completely unbiased classification models because they ultimately always make implicit (but unobjective) claims to represent their subjects. These issues prompt different modes of tactical engagement with the roles of CV in biometric classification and taxonomic imaging.

For example, Jake Elwes' video *Machine Learning Porn* (2016)[62] plays with moral prejudices as well as perceptive and technical flaws that influence AI design within the context of online content "propriety." Elwes took the open_nsfw convolutional neural network trained on Yahoo's model for detecting "sexually explicit" or "offensive" content and repurposed its recognition classifiers as parameters to generate new sequences of abstract images with a "porny" allusiveness. This work encapsulates the problem of AI-powered Internet censorship (e.g., through content filtering on social media), which has wide-reaching consequences for the freedom of creative expression and the overall variety and character of cultural production.[63] Another Elwes' generative video, *Closed Loop* (2017),[64] keeps the CV biometry algorithms "at home" by establishing a loop between a mutually feeding text-to-image and image-to-text model, whose inaccuracies and biases imply the applied AI's ethical issues in a wittily unpredictable continuum.[8]

Shinseungback Kimyonghun's installation *Mind* (2019)[65] uses FER+ Emotion Recognition annotation and ONNX Runtime scoring engine for the emotion analysis of the last hundred gallery visitors' facial expressions to drive fifteen ocean drums that generate an impressive minimalist soundscape with an overhead camera as a single indicator of the machinic gaze (Figure 7).

Some works, such as Zach Blas' *Facial Weaponization Suite* (2011 – 2014) and *Face Cages* (2015 – 2016), aim to protect privacy with practical solutions that obstruct facial recognition algorithms, while others, such as Avital Meshi's *Classification Cube* (2019), Adam Harvey and Jules LaPlace's *Exposing.ai* (2021), or Coralie Vogelaar's *Random String of Emotions* (2018), provide estheticized deconstructions of automated emotion analysis. For instance, in *Random String of Emotions* (2018),[47] Vogelaar's software generated a set of her portraits based on a random chain of temporal visual segments that constitute a facial expression. An emotion recognition program then analyzed the portraits and assigned them the percentages for six basic emotions: happy, sad, angry, surprised, scared, and disgusted. The resulting book presents all tagged portraits whose complex but non-existing emotional configurations underline the mechanistic nature and implicit absurdities of statistical models used for expression recognition and classification.

Rafael Lozano-Hemmer's commemorative work *Level of Confidence* (2015)[66] emphasizes the somber aspects of facial recognition. This interactive setup matches visitor's faces with the portraits of 43 students from the Ayotzinapa Normalista school in Iguala, Mexico, who were kidnapped and most likely mass-murdered and incinerated in an incident on September 26, 2014, which involved the Mexican government, police forces, and drug cartels. When the visitor steps in front of a screen with an overhead camera, the CV software tracks their face and finds the image of a student with the best matching facial features. The student's image is displayed together with the matching accuracy or the "level of confidence." Lozano-Hemmer intentionally used the same biometric surveillance algorithms that the military and police typically use to look for suspects: Eigen, Fisher, and LBPH.

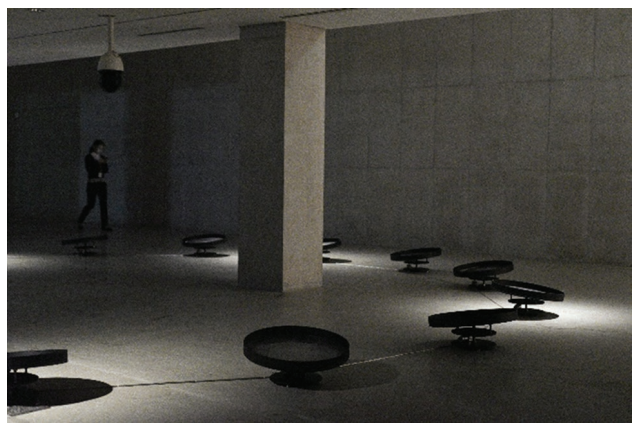

**Figure 7.** Shinseungback Kimyonghun, *Mind* (2010). Installation view. Photograph: Shinseungback Kimyonghun. Courtesy of the artists

Comparably, Mushon Zer-Aviv's *The Normalizing Machine* (since 2018)[67] places its recursive critique of normative statistics and automated biometric classification within a historical context (Figure 8). This installation's visitors face a serial line-up of pairs of previously recorded visitors and point out the one that looks more "normal." Their portraits, captured during this process, are added to the training dataset, and their selection decisions modify a generative model that continuously visualizes the facial aggregate of "normalcy" in a separate image. *The Normalizing Machine* adopts the methodology and esthetics of Alphonse Bertillon, the French forensics pioneer who developed a system for standardizing, indexing, and categorizing human faces in the late 1800s. The installation setup also includes references to Alan Turing, who arguably hoped that AI would transcend the systemic bias that criminalized his deflection from the sexual norm.[9]

In influencing and determining our ways of seeing and decision-making, modern AI relies on data retrospection

---

8    This work's elegant concept and methodology, which Elwes reused in *A.I. Interprets A.I.: Interpreting "Against Interpretation" (Sontag 1966)* (2023), were cloned by Theodoros Papatheodorou and Jack DiLaure's *Visual Dialogues* (2023), which invites the visitor to submit a hand drawing to initiate the loop between the prompt-based CLIP for image-to-text and Stable Diffusion for text-to-image conversion.

9    After being celebrated for his contributions to Britain's cryptographic counterintelligence in World War II, Turing was arrested for homosexual activity in 1952. To evade the prison sentence, he chose to undergo estrogen therapy which caused him to develop female bodily characteristics, become depressive, and allegedly commit suicide in 1954.

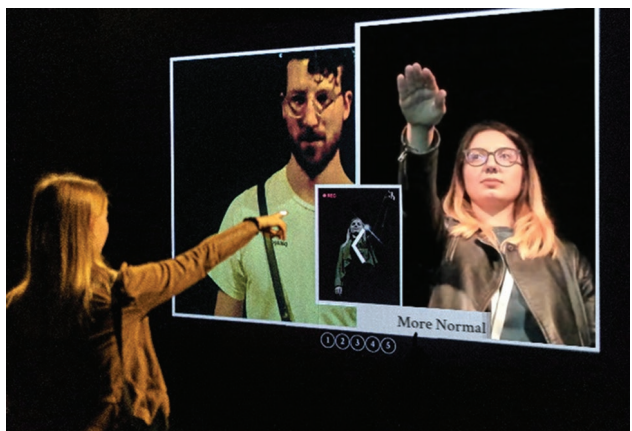

**Figure 8.** Mushon Zer-Aviv, *The Normalizing Machine* (since 2018). Installation view at the Fotomuseum Winterthur, 2019. Photograph: Lucidia Grande. Courtesy of the artist

and lacks the capacities for commonsense reasoning and inference, which makes it inflexible and limits its range of meaningful and responsible applications.[10,11] Artworks such as *Machine Learning Porn*, *Level of Confidence*, and *The Normalizing Machine* point to the fact that CV errors combine technical deficiencies in recognition range or accuracy with much more decisive human factors, such as cognitive flaws, prejudices, biases, and conflicting economic or political interests.

## 2.4. Ethical and epistemic limits

Human factors impact AI development, extensive industrialization, and sometimes rushed application in sensitive areas, such as jurisdiction, HR, insurance, or health care by retaining or amplifying the existing cultural, economic, linguistic, ethnic, gender, and other inequities.[4,68-70] However, many undesirable human-induced byproducts, such as biases, remain unanticipated during research or unregistered in testing. Instead, they are often mitigated after being detected in the deployed AI products, which hint at the soundness of the safety culture in AI engineering and the social responsibility standards of the AI industry.

These issues are particularly conspicuous in face detection and identification due to the facial convergence of evolutionarily significant visual markers and the psychological role of the face as a representative locus of the self and identity. Flaws of network architectures used for facial recognition and biases in facial data annotation and classification have been identified by both scientists[71] and artists, as exemplified by Joy Buolamwini and Timnit Gebru's *Gender Shades* (2018).[72] It started in 2017 as scientific research for Buolamwini's master thesis and morphed into a documentary and educational project with artistic overtones. Using a custom benchmark dataset of diverse skin types based on 1,270 images of parliamentarians from three African and three European countries, Buolamwini and Gebru assessed the accuracy of several corporate facial classifiers (Adience, IBM, Microsoft, and Face++) concerning gender, skin type, and gender/skin type intersection. They showed that the error rate of the tested classifiers was significantly higher for women with darker skin color and published their dataset to be used for accuracy calibration. Their findings gained public attention and influenced United States (US) policymakers and the AI industry.[73]

Kate Crawford and Trevor Paglen's multipart project *Training Humans* (2019 – 2020)[74] followed a similar agenda. Its critique of the racial bias manifest in CV training datasets and the use of facial images and videos without consent for building these datasets was widely credited with raising public attention about the problems in the online database ImageNet, whose more than 14 million Internet-scraped pictures have been used in ML since 2009. However, claims[75] that part of this project, the work called *ImageNet Roulette* (2019), stirred ImageNet to excise 600,000 offensive synsets are purely conjectural.[10] More importantly, it was revealed that the creators of JAFFE, CK, and FERET datasets (featured alongside ImageNet in *Training Humans*) had duly obtained permissions from the depicted persons, whereas Paglen and Crawford themselves collected, reproduced, and exhibited images from these datasets without consent, and made technical errors in their critical analysis of the purpose of several datasets. It is no less dubious that Paglen and Crawford found it appropriate to partner with the high fashion industry (Prada Mode Paris) to promote *Training Humans*, somehow overlooking its forefront position in the sustainability and environmental crises and its baggage of exploitative business practices.[78,79] Perhaps the critical compromises and ethical inconsistencies of this project may be recognized as tradeoffs of Paglen's position in the mainstream art world.[80]

A slew of artworks centers on human perceptive flaws that slip into perceptive apparatuses. For instance, Benedikt Groß and Joey Lee's online project *Aerial Bold* (since 2016)[81] uses CV pareidolia[11] to make a generative typeface from alphabet shapes found in aerial imagery. It features detailed documentation, a font catalog, and an interactive word processor where visitors can enter text and choose the font size, line spacing, different font classes, and source locations for typesetting. Related works include Shinseungback Kimyonghun's *Cloud Face* (2012), Onformative Studio's *Google Faces* (2013), and Driessens and Verstappen's *Pareidolia* (2019). Although they prioritize the decorative and amusing aspects of CV over its critique, these works hint at the shadier sides of errors in automated vision systems.

---

*10   ImageNet's staff had already begun addressing its problems in 2018,[76] and their statement about the database improvements,[77] sourced in several writings about the ImageNet Roulette, makes no mention of that project and no reference to the public criticism stirred by the art scene as the motives for removing the synsets. Synsets are the groupings of synonymous words that express the same concept. They are used in the NLP modules of CV architectures to generate image tags or descriptions.*

Unlike the biases in AI-powered CV, the risky "creative biases" and idiosyncrasies are desirable in tactical art, where they can catalyze conceptual cogency and expressive economy. For instance, Jennifer Gradecki and Derek Curry's *Boogaloo Bias* (2021)[83] forsakes Lozano-Hemmer's solemn tone in *Level of Confidence* (2015, discussed in the previous section)[66] to cast a sarcastic look at the biases and errors in CV translation processes and the impact of datasets and accuracy thresholds on false positives in police surveillance and arrest policies (Figure 9). Trained on the actors' faces from the movie *Breakin' 2: Electric Boogaloo* (1984, directed by Sam Firstenberg), the facial recognition algorithm in this work brute-forces the generation of leads to "identify" members of the Boogaloo Bois anti-law enforcement militia in their live video feeds and protest footage. "Brute-forcing" in ML facial recognition adopts the methods of police investigations in the US, which occasionally compensate for the absence of high-quality suspects' images by substituting artist sketches with filtered social media photographs, computer composites, or celebrity images that resemble suspects. Hence, unlike regular facial recognition algorithms, tuned to minimize the matching uncertainty level, *Boogaloo Bias* is optimized to return the highest number of suspects.

In several landmark projects bolstered by neat and striking humor, Sebastian Schmieg expands the problem space of image recognition accuracy and normalization by emphasizing the fundamental but insufficiently investigated philosophical dimensions of AI ontology and epistemology.[84] He introduces deliberately reduced and unconventional (seemingly absurd) taxonomies into the CV classification libraries used in interactive setups to process visitors' uploaded or captured pictures. For example, online visitors of the *Decisive Camera* (2017 – 2018)[85] can upload an image that will be classified within a taxonomic space of only four categories (Problem, Solution, Past, and Future) with a probability percentage for each category. The classification dataset was created in the project's initial phase, which invited visitors to assign one of these categories to the images in the archive of the Photographers Gallery where the work was later installed. Similarly, the *Decisive Mirror* (2019)[86] analyzes and classifies visitors' faces based on unconventional "traits" (predefined classification categories), such as "aliveness," "imaginariness," or "one-of-themness," to underline the arbitrariness and randomness that creep into the ML profiling categories (Figure 10). It is worth noting that in these projects, Schmieg already addressed the experiential revelation of AI's limitations and flaws to challenge the presumptive objectivity and interpretive validity of computational processes[87] – the same poetic punctum that garnered acclaim for Paglen and Crawford's latter *ImageNet Roulette* (2019, discussed above).[74]

The statistical nature of CV and the related ambiguities of ML are sometimes utilized to address the questionable intersections of AI and creativity. In Adam Basanta's installation, *All We'd Ever Need Is One Another* (2018),[88] custom software randomizes the settings of two mutually facing flatbed scanners so that in every scanning cycle, each captures a slightly altered mix of the facing scanner's light and its own unfocused scanning light reflected off the facing scanner's glass plate. The perceptual hashing algorithms use parameters such as aspect ratio, composition, shape, and color distribution to compare each new scan to an image database scraped from freely accessible online artwork repositories. When the comparison value between the scan and the most similar database image exceeds 83%, the software declares a "match," selects the scan for printing, and labels it according to the database image metadata. After it printed one of the scans and labeled it "85.81%_match: Amel Chamandy 'Your World without Paper,' 2009," Canadian artist Amel Chamandy initiated a legal action over the intellectual property rights against Basanta because of the reference to her photograph. However, the "85.81%_match" is not for sale, and Basanta does not use it for direct commercial gains by any other means. He consistently applied the functional logic of ML to leverage the open-endedness of creative work and disturb the entrenched notions of agency, authorship, originality, and intellectual property crystalized in copyright laws.[89]

This line of practices places the conceptual, technical, and broader sociopolitical problems of AI engineering and application firmly within the human context, even in cases when their expressive cogency and professional ethics are debatable. They point to the reflections of human nature and the (political) rationales encoded into CV software and hardware architectures, which modulate both the power vectors of corporate/state AI and the poetic flavors of artistic interventions that criticize them.

---

11   Pareidolia is a tendency to recognize patterns within random visual data.[82]

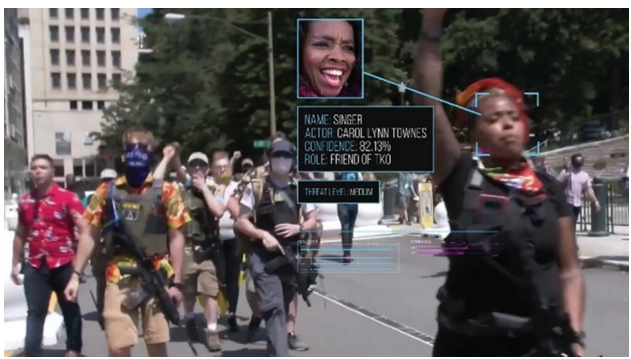

**Figure 9.** Jennifer Gradecki and Derek Curry, *Boogaloo Bias* (2021). Screenshot of Boogaloo Bias facial recognition system analyzing news footage of an anti-gun control protest. Courtesy of the artists

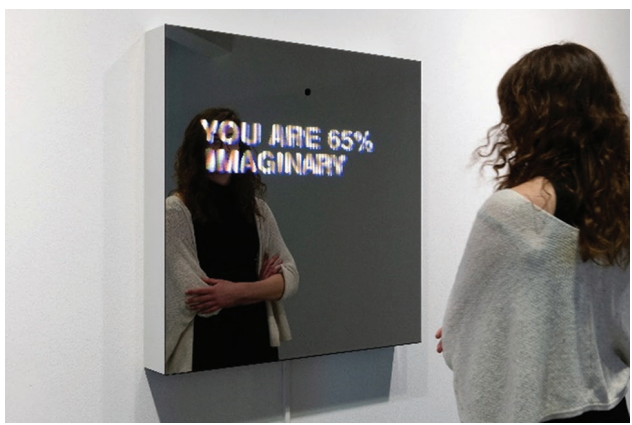

**Figure 10.** Sebastian Schmieg, *Decisive Mirror* (2019). Installation view photograph. Courtesy of the artist

## 3. Discussion

Critical art practices with CV are a valuable contribution to AI art. They reveal a range of AI's CV-related issues, such as the imposed normalization of behavior and the influence of biometric profiling on social media, democratic processes, political convictions, self-determination, and privacy. They highlight the roles of AI-powered surveillance in questionable forms of societal control, such as manipulating public presence, self-perception, labor maximization, and profit extraction. By emphasizing the differences between computer and human vision, these practices show how the deployment of automation in situations where human intuition and inference are norms changes the epistemological and ethical foundations for decision-making. They stress that CV is part of a higher algorithmic order riddled with technical imperfections, inadequacies of statistical modeling and prediction, human biases and prejudices, and conflicting economic and political interests.

By underlining the misalignment between the myths about digital technologies and their actual implementation, they undermine positivist worldviews that equate quantification and calculation with objective truth, challenge the authority of the technocratic elite, and establish alternative narratives to corporate techno-solutionism.[7,90] They nourish the necessary macro-level skepticism about pervasive technologies, such as AI, informed by their assessment from broader historical, economic, and social perspectives.[91] While other tactical (AI) art practices make equivalent critical points, the cognitive power of vision amplifies their persuasiveness in CV-related works.

At the same time, CV-related critical artworks' limited capabilities for making palpable social changes beyond academic or popular discourse represent a larger set of conceptual and methodological challenges across tactical AI art, tactical media art, and computational art.

### 3.1. Challenges

Tactical AI art continues computational art's historical tendency toward technocentrism, initially imposed by high cognitive demands of computer technology, which tacitly conflates artmaking with the skillful handling of creative devices and often incentivizes artists toward technical virtuosity devoid of critical distance or playful irreverence.[92,93] This techno-fetishist mentality often reinforces a naïve lack of understanding that the poetic sense of production techniques in the arts is fundamentally defined by conceptual thinking and meaningful contextualization.

As in computational art, expressive similarities are frequent in tactical AI art. While some artworks leverage the well-sanctioned referencing of other artworks as an asset, in cases where obvious or easily discoverable similarities remain unacknowledged, they become liabilities that raise concerns about their creators' art-historical literacy or professional ethics.[94] Similar to tactical media art, the actual extent and longer-term social benefits of many works (even those initially recognized for making palpable changes) turn out to be debatable or have been contested. Pertinent and well-conceived critical ideas are sometimes rendered as unengaging, vague, ineffective, or counter-effective works, represented in overly didactic setups, or supported by inflated theoretical rhetoric.[95,96] Tactical AI artworks are largely consumed and praised by academic communities prone to developing echo chambers, while their real-world critical viability requires unbiased assessment. The general audience, which is supposed to be central to tactical art, can easily recognize the academization of tactical values or purposes as aloofness or cynicism, leading to indifference, distrust, or resentment. The cultural value of tactical AI art practices depends on their adequate positioning, representation, and preservation, whose range of problems has not been solved despite extensive efforts.[97]

The most elusive source of tactical AI art's issues is its sociotechnical milieu, which simultaneously provides critical affordances and threatens to relativize, debase, neutralize, or instrumentalize their outcomes. Since the mid-2010s, the AI industry has been sponsoring and promoting AI art as a powerful marketing and public relations instrument, and the mainstream art market has absorbed it in its trading portfolio. Thus, many decorative non-tactical artworks serve as spectacles of corporate AI power,[98] while tactical ones frequently get recuperated, occasionally become mere reflections of their targets,[99] or lapse into the mystification of technocracy whereby a class of tech-savvy artisans acts on behalf of "the (lay) people" by articulating a vision of individual freedom realizable from within the power structures of the information society.[100] The corporate sector leverages AI art's sociotechnical entanglement to systematically assimilate and often exploit activist practices for refining the normalization or circumventing problems they identified instead of correcting them.[101] For instance, works that unveil the vulnerabilities of face recognition algorithms can stir up their technical improvements under the existing application logic. For all these reasons, even when proactively intervening, tactical AI art opens questions beyond its apparent contributions.[102] Should an activist action end up (directly or indirectly) being used by the AI industry to enhance its profitable instruments or remedy its public image without necessarily improving its techno-ethical standards? Could tactical art disrupt the corporate AI regimen with lasting and desirable social consequences, and how effectively can it incite or enhance government policies for accountability and regulation of private AI businesses with global influence?

Tactical AI art shares these political vulnerabilities with its audience. Modern AI's social impact is marked by the uneven performance of applied AI systems: in some contexts, they are beneficial; in others, superfluous, absurd, abusive, or dangerous.[103] The causes of this performative unevenness are not purely technical. They relate to a shady facet of AI tech's commercial success, which concerns substantial investments in, yielding profits from, and normalizing the technologies for data capture, analysis, and monetization. The vast asymmetry between societal cost and private gain in the inflated rollout of these technologies aids the neoliberal political backlash that de facto seeks to bypass or erode civil, labor, and human rights.[104] However, these processes are substantiated not only by the AI industrialists and investors but also by most users willing to trade their civic duties for the benefits of surveillance-based consumerism, comfort, or security. The implicit public complacency reinforces the inequalities embedded in and served by automation and drives society toward a risky reliance on brittle data-based classification/prediction systems.[8,10]

Tactical AI art's main challenges lurk in its theoretical foundations. As Martin Zeilinger noted,[105] media art activism has been largely influenced by a media theory canon from the left, with authors such as Rita Raley,[6] Alexander Galloway,[106] McKenzie Wark,[107] or Critical Engineering Working Group.[108] This canon draws upon Michel de Certeau's[109] conceptual distinction between strategy and tactics in cultural contexts. De Certeau defines strategy as a goal-oriented set of practices focused on instrumentalizing the structural affordances of their environmental, social, or technological substrates. The strategy serves administrative agendas by drawing on system-inherent control architectures, often to contain divergent elements, for instance, through algorithmic labor or consumer management systems. Conversely, tactical practices articulate anti-authoritarian and oppositional responses to strategically advantageous positions of dominance through actions that are responsive, fluid, and embedded in the systems they challenge. However, theoretical notions about the strategic vs. tactical dichotomy and the suitability and meaningful impact of digital activist operations within the systems and by means of strategic power seem to be too vague, permissive, compromising, or implicitly compliant when facing the extent and sophistication of modern AI's techno-economic regime and its broader realm of neoliberal info-capitalism.[110] Specifically, they imply the question: what compromises, vulnerabilities, and other trade-offs come with deciding or accepting to operate critically within (not against) the infrastructures, protocols, and ultimately the rules of that realm? Many tactical artworks, even if praised by academia or the art scene, fail to instigate noticeable long-term changes primarily because their makers could not reach the optimal ratio between defiance, bravery, cunning, ambition, and survival instinct in answering that question. Under present circumstances, it seems that such conceptualization of art activism is often more effective in unwillingly aiding the strategic power to simultaneously exploit (recuperate) and marginalize critical artworks than in allowing them to disrupt that power. There is no denying that successful critical AI artworks are effective in raising public awareness about AI issues. They usually do it with more panache and charm than investigative journalism or other non-artistic forms of activism, but they mostly remain within that "signal layer" and are more detectable as a discursive matter, particularly in art academic circles, than by the outcomes of social or political actions in their wake.

Hence, tactical AI art's conceptual models, underlying motivations, and sociotechnical realities coexist in a contradiction whose ethical aspects deserve more scrutiny. This dissonance provides ground for an argument that artists may be concerned with criticality more because they were taught that being critical is good than because of their experientially informed need to address specific sociopolitical concerns. The result is artmaking in which criticality figures more as a floating rhetoric than as a device for political action. It perhaps also explains why artists who work with AI still seldomly address the implicit complacency toward AI power regimens that AI art shares with its audience. On the other hand, the shrewdness of many critical artworks indicates that the tactical dissonance does not persist due to the lack of artists' creative wit but probably due to a combination of the conceptual underpinnings of media art activism, the contemporary crisis of political opinion, and the deficit of political ideas in developed post-industrial democracies.[111] However, even if tactical art's contributive scope remains constrained to mediating the discomforting symptoms of modern society and occasionally inciting their remedies, it can be improved.

### 3.2. Prospects

The versatility and ubiquity of digital computing and AI, as one of its flourishing application domains, are structural means for power concentration and administrative control. Such governing instruments should be chiefly in the hands of the citizenry, encompassing majorities and minorities, communities, and individuals. Thus, critical AI art needs to question whether and how the forms, functions, and availability of the computational/AI infrastructure can be brought under transparent and humane democratic control. By identifying, acknowledging, and understanding the field's issues outlined above, artists can find new ways not just to pose such questions but also to take part in answering them. The import of human vision puts artists who critique its computational emulations in a good position to anticipate, articulate, and foster changes that lead to more just and wiser governing of AI's sociopolitical and economic regimens. These improvements would benefit all areas of AI art and its superset computational art in establishing a more poetically cogent, socially responsible, and politically effective stratum.

To challenge or tangibly affect AI power and mobilize the audience with a lasting impact, artists need to maintain a sharp outlook on their poetic devices and balance procedural skills with motivational sincerity and ideational cogency. This ethos of balanced competencies requires cultivation through experimental freedom, playfulness, conceptually strong hacking, and imaginative discovery. Artists can leverage epistemic humility to develop more rigorous criteria for creative thinking and acquire wider knowledge of the historical, theoretical, cultural, and political contexts in which they produce and present their works.[112] Developing this rigorous approach will help them address potentially adverse expressive circumstances and clear the way for strong ideas in devising novel ways to harness the physiological and intuitive power of intersecting humans with machine vision. The consequentiality of CV and AI technologies obliges artists not just to exploit them as expressive means but to recognize, deconstruct, and explore the injustices in the notional, relational, economic, and other layers of their application. By combining sincerity in assessing the sociotechnical conditions of their everyday and creative lives with a deeper understanding of AI issues and stakeholders, artists can overcome the (implicit) cynicism or unconscious resignation that undermines their criticality.

To counter the recuperative pull of the corporate AI sector, art market, and academia, artists should strive for integrity by identifying and correcting the systemic value issues in these domains. Artists' proverbial inclination toward opportunism calls for decidedly independent professional strategies and sophisticated interventions that demask and evade the lures of commodification and complacency. These strategies require viewing art and technology as dispositives within anthropological and sociocultural perspectives and recognizing (and remembering) that creativity is driven by competitive ambitions and is thus inherently instrumentalizable. This recognition informs the ability to resist prioritizing career building over artmaking, pursue external support with skepticism toward institutional rationales for art sponsorship, and be open to taking genuine risks by evolving potentially hazardous ideas.

The political nature of AI technologies[113] largely remains concealed behind the big business glamorization of their successes. To reveal it, critical artworks can go beyond rhetorical questions to disclose who makes the decisions about AI's application domains and optimization goals, who benefits from them, and how. Their impact can be improved by bolder probing and deeper problematizing modern AI's underlining concepts such as intelligence, creativity, expressive agency, authorship, intellectual labor, ownership, authenticity, accuracy, and fairness.[114] The flexibility and mutability of these concepts are inherent to sociocultural dynamics, and technologies such as CV can be used to reconfigure or extend them in insightful ways. By demystifying the expressive contexts and the seemingly radical capabilities of their tools through experiences that engage the audience's acumen and imagination, artists can leverage AI issues as assets for gaining new insights about human nature and relations.

## 4. Conclusion

The contemporary AI-infused sociotechnical landscape provides a range of resources for artists to reveal its paradoxes, flaws, and injustices and points out that science, technology, businesses, and education need a thorough reconfiguration and improvement of epistemological and ethical standards facing the increasing complexity of human existence. Given these potentials and stakes, a constructive appraisal of critical AI art's transformative values should reassess the viability of the prevalent theoretical premises, such as the tactical versus strategic distinction in cultural practices and recognize the adjective "tactical" as overly hedging under present sociotechnical circumstances, particularly by the assumption of tactical practices' fleeting nature. The claim that tactical art practices do not necessarily have to provoke political action and may be justified by the intellectual stimulation they provide is undermined by the ethically charged critique in their poetic core, which makes a strong intuitive drive to take them as political acts that promote or at least imply a call to action. Political actions can yield longer-term changes only if adopted widely enough or sufficiently sustained.

Matching the complexities and ambiguities of contemporary AI's political reality entails building an authentic and robust conceptual framework for critical artmaking instead of merely importing models from disciplines such as sociology, cultural theory, or philosophy. This framework requires more stringent self-reflexivity and readiness to derive sober insights about pragmaticism and resilience from the strategic realms it aims to disrupt, which may result in artists' more radical and equally responsible approaches to risk-taking and provocation. Crucially, its affordances should be informed by seeking and understanding the accounts in which the means (technologies) developed to serve the long-term interests of the political power have been converted into serving the long-term interests of its opposition. This paper contributes to building such a framework by providing a starting point for a systematic study of CV-related critical art.

## Acknowledgments

None.

## Funding

## Conflict of interest

## Author contributions

## Ethics approval and consent to participate

## Consent for publication

## Availability of data

## Further disclosure

(i) This article includes several topically relevant passages from the paper *The Mechanical Turkess: Tactical Media Art and the Critique of Corporate AI* (https://arxiv.org/pdf/2402.17490) by the same author, which is not intended for republication.

(ii) The artists provided all figures in the article and gave their permission for publication. The photograph/image credits in figure captions are as specified by the artists. No figures were taken from published articles.